\documentclass[12pt,preprint]{aastex}

\slugcomment{A slightly edited version of this article
appeared in the 9 October issue of Nature (2003).  The Nature version 
is available from http://www.nature.com}

\shorttitle{CMB Data suggest Dodecahedral space topology}
\shortauthors{Luminet et al.}

\begin{document}

\title{Dodecahedral space topology
as an explanation for weak
wide-angle temperature correlations
in the cosmic microwave background}

\author{J.-P. Luminet}
\affil{Laboratoire Univers et Th\'eories, CNRS-UMR 8102, Observatoire 
de Paris, \\ F-92195 Meudon Cedex (France)}
\email{jean-pierre.luminet@obspm.fr}

\author{J. Weeks}
\affil{15 Farmer St., Canton NY 13617-1120 
(USA)}
\email{weeks@northnet.org}

\author{A. Riazuelo}
\affil{Service de Physique Th\'eorique, CEA/DSM/SPhT, CEA/Saclay, 
\\ F-91191 Gif-sur-Yvette Cedex (France)}
\email{riazuelo@spht.saclay.cea.fr}

\author{R. Lehoucq}
\affil{CE-Saclay, DSM/DAPNIA/Service 
d'Astrophysique, \\ F-91191 Gif sur Yvette Cedex (France)}
\email{lehoucq@cea.fr}

\and

\author{J.-P. Uzan}
\affil{Laboratoire de Physique Th\'eorique, 
CNRS-UMR 8627, B\^at. 210, Universit\'e Paris XI,
\\ F-91405 Orsay Cedex (France)}
\email{uzan@iap.fr}

\begin{abstract}
Cosmology's standard model posits an infinite flat universe forever
expanding under the pressure of dark energy.  
First-year data from the Wilkinson Microwave Anisotropy Probe (WMAP) 
confirm this model to spectacular precision on all but the largest 
scales (Bennett {\it et al.}, 2003 ; Spergel {\it et al.}, 2003).
Temperature correlations across the microwave sky match expectations on scales 
narrower than $60^{\circ}$, yet vanish on scales wider than $60^{\circ}$. 
Researchers are now seeking an explanation of the missing wide-angle 
correlations (Contaldi {\it et al.}, 2003 ; Cline {\it et al.}, 2003).  
One natural approach questions the underlying geometry of space, namely 
its curvature (Efstathiou, 2003) and its topology (Tegmark {\it et al.}, 2003).  
In an infinite flat space, waves from the big bang would fill the universe 
on all length scales.  The observed lack of temperature correlations on scales
beyond $60^{\circ}$ means the broadest waves are missing, perhaps because space itself
is not big enough to support them.  

Here we present a simple geometrical model of
a finite, positively curved space -- the Poincar\'e dodecahedral space -- which
accounts for WMAP's observations with no fine-tuning required.  
Circle searching (Cornish, Spergel and Starkman, 1998) may confirm the model's topological predictions, 
while upcoming Planck Surveyor data may confirm its predicted density
of $\Omega_0 \simeq 1.013 > 1$.  If confirmed, the model will answer the
ancient question of whether space is finite or infinite,
while retaining the standard Friedmann-Lema\^\i{}tre foundation for local physics.

\end{abstract}

\keywords{cosmology: theory, large scale structure: topology}

\section{CMB Data Suggest Spherical Space with Dodecahedral Topology}

Temperature fluctuations on the microwave sky may be expressed as a sum
of spherical harmonics, just as music and other sounds may be 
expressed as a sum of ordinary harmonics.  A musical note is the sum
of a fundamental, a second harmonic, a third harmonic, and so on. The relative strengths
of the harmonics -- the note's spectrum -- determines the tone quality, 
distinguishing, say, a sustained middle C played on a flute from the same
note played on a clarinet.  Analogously, the temperature map on the microwave sky
is the sum of spherical harmonics.  The relative strengths of the harmonics
-- the power spectrum -- is a signature of the physics and geometry of the universe.
Indeed the power spectrum is the primary tool researchers use to test 
their models' predictions against observed reality.

The infinite universe model gets into trouble at the low end of the power spectrum
(Figure 1).  The lowest harmonic -- the dipole, with wave number $\ell = 
1$ -- is unobservable because the Doppler effect of the solar system's
motion through space creates a dipole a hundred times stronger, 
swamping out the underlying cosmological dipole. The first observable harmonic
is the quadrupole, with wave number $\ell = 2$. WMAP found a quadrupole only about 
$1/7$ as strong as would be expected in an infinite flat space.  
The probability that this could happen by mere chance has been estimated 
at about a fifth of one percent (Spergel {\it et al.}, 2003). 
The octopole term, with wave number $\ell = 3$, is also weak at $72 \%$
of the expected value, but not nearly so dramatic or significant as the quadrupole.
For large values of  $\ell$, ranging up to $\ell = 900$ and corresponding to
small-scale temperature fluctuations, the spectrum tracks the infinite universe 
predictions exceedingly well.

Cosmologists thus face the challenge of finding a model that accounts
for the weak quadrupole while maintaining the success of the infinite flat 
universe model on small scales (high $\ell$).  The weak wide-angle temperature
correlations discussed in the introductory paragraph correspond directly
to the weak quadrupole.

Microwave background temperature fluctuations arise primarily (but not exclusively)
from density fluctuations in the early universe, because photons travelling
from denser regions do a little extra work against gravity and therefore arrive
cooler, while photons from less dense regions do less work against gravity
and arrive warmer.  The density fluctuations across space split into a sum
of 3-dimensional harmonics -- in effect the vibrational overtones of space itself --
just as temperature fluctuations on the sky split into a sum of 2-dimensional
spherical harmonics and a musical note splits into a sum of 1-dimensional harmonics.
The low quadrupole implies a cut-off on the wavelengths of the 3-dimensional harmonics. 
Such a cut-off presents an awkward problem in infinite flat space, because 
it defines a preferred length scale in an otherwise scale-invariant space.
A more natural explanation invokes a finite universe, where the size of space
itself imposes a cut-off on the wavelengths (Figure 2).  Just as the vibrations
of a bell cannot be larger than the bell itself, the density fluctuations in space
cannot be larger than space itself.  While most potential spatial topologies
fail to fit the WMAP results, the Poincar\'e dodecahedral space fits them strikingly well.

The Poincar\'e dodecahedral space is a dodecahedral block of space with opposite
faces abstractly glued together, so objects passing out of the dodecahedron
across any face return from the opposite face.  Light travels across the faces
in the same way, so if we sit inside the dodecahedron and look outward across
a face, our line of sight re-enters the dodecahedron from the opposite face. 
We have the illusion of looking into an adjacent copy of the dodecahedron. 
If we take the original dodecahedral block of space not as a Euclidean 
dodecahedron (with edge angles $\simeq 117^{\circ}$) but as a spherical dodecahedron 
(with edge angles exactly $120^{\circ}$), then adjacent images of the dodecahedron fit 
together snugly to tile the hypersphere (Figure 3b), analogously to the way adjacent
images of spherical pentagons (with perfect $120^{\circ}$ angles) fit snugly to tile
an ordinary sphere (Figure 3a). Thus the Poincar\'e space is 
a positively curved space, with a multiply connected topology whose volume 
is 120 times
smaller than that of the simply connected hypersphere (for a given curvature radius).

The Poincar\'e dodecahedral space's power spectrum depends strongly on the 
assumed mass-energy density parameter $\Omega_0$ (Figure 4). The octopole term
($\ell = 3$) matches WMAP's octopole best when $1.010 < \Omega_0 < 1.014$. 
Encouragingly, in the subinterval $1.012 < \Omega_0 < 1.014$  the quadrupole ($\ell  = 2$)
also matches the WMAP value. More encouragingly still, this subinterval agrees
well with observations, falling comfortably within WMAP's best fit range of
$\Omega_0 = 1.02\pm0.02$ (Bennett {\it et al.}, 2003).

The excellent agreement with WMAP's results is all the more striking because
the Poincar\'e dodecahedral space offers no free parameters in its construction.
The Poincar\'e space is rigid, meaning that geometrical considerations require
a completely regular dodecahedron.  By contrast, a 3-torus, which is nominally
made by gluing opposite faces of a cube but may be freely deformed to any 
parallelepiped, has six degrees of freedom in its geometrical construction. 
Furthermore, the Poincar\'e space is globally homogeneous, meaning that its 
geometry Ñ- and therefore its power spectrum Ñ- looks statistically the same
to all observers within it.  By contrast a typical finite space looks different
to observers sitting at different locations.

Confirmation of a positively curved universe ($\Omega_0 > 1$) would require revisions
to current theories of inflation, but the jury is still out on how severe 
those changes would be.  Some researchers argue that positive curvature 
would not disrupt the overall mechanism and effects of inflation, but 
only limit the factor by which space expands during the inflationary 
epoch to about a factor of ten (Uzan, Kirchner and Ellis, 2003).  Others claim that 
such models require 
fine-tuning and are less natural than the infinite flat space 
model (Linde, 2003).

Having accounted for the weak observed quadrupole, the Poincar\'e dodecahedral 
space will face two more experimental tests in the next few years:

\begin{itemize}
    
    \item 
    The Cornish-Spergel-Starkman circles--in--the--sky method 
(Cornish, Spergel and Starkman, 1998) predicts temperature correlations along matching circles
in small multiconnected spaces such as this one. 
When $\Omega_0 \simeq 1.013$ the horizon radius is about 0.38 in units of 
the curvature radius, while the dodecahedron's inradius and outradius 
are 0.31 and 0.39, respectively, in the same units; as a result, the 
volume of the physical space is only $83 \%$ the volume of the horizon sphere.
In this case the 
horizon sphere self-intersects in six pairs of circles of angular radius
about $35^{\circ}$, making the dodecahedral space a good candidate 
for circle detection if technical problems (galactic foreground removal,
integrated Sachs-Wolfe effect, Doppler effect of plasma motion) can be overcome.
Indeed the Poincar\'e dodecahedral space makes circle searching easier than 
in the general case, because the six pairs of matching circles must a priori
lie in a symmetrical pattern like the faces of a dodecahedron, thus allowing 
the searcher to slightly relax the noise tolerances without increasing the
danger of a false positive.
    
    \item
    
   The Poincar\'e dodecahedral space predicts $\Omega_0 \simeq 1.013 > 1$.  
   The upcoming Planck surveyor data (or possibly even the existing WMAP data 
   in conjunction with other data sets) should determine $\Omega_0$ to within $1 \%$.
   Finding $\Omega_0 < 1.01$ would refute the Poincar\'e space as a cosmological model,
   while $\Omega_0 > 1.01$ would provide strong evidence in its favour.

\end{itemize}

\section{Conclusion}

Since antiquity humans have wondered whether our universe is finite or infinite.
For most of the past two millennia Europeans held the Aristotelian view of the 
universe as a finite ball with a spherical boundary.  The invention of the telescope
in 1608 revealed the universe to be much larger than Aristotle had imagined.  
Thus even though Galileo and Kepler clung to AristotleÕs model, their successors
Bruno, Descartes and especially Newton embraced the idea of infinite space.  
Nevertheless, some scientists were as uncomfortable with an infinite universe
as they were with Aristotle's hypothetical boundary.  In 1854 Georg Riemann cut 
the Gordian knot by proposing the hypersphere as a model of a finite universe with
no troublesome boundary.  By 1890 Felix Klein discovered the more general concept 
of a multiconnected space (recall Figure 2) and during the early years of the 
twentieth century Einstein and others preferred finite universe models.  
Nevertheless, by the 1930's the vast size of the observable universe had become
known and the pendulum swung back towards infinite models.  Now, after more than
two millennia of speculation, observational data might finally settle this ancient
question once and for all.

\acknowledgments

J.R.W. thanks the MacArthur Foundation for its generous support.

\clearpage


\begin{figure}
\plotone{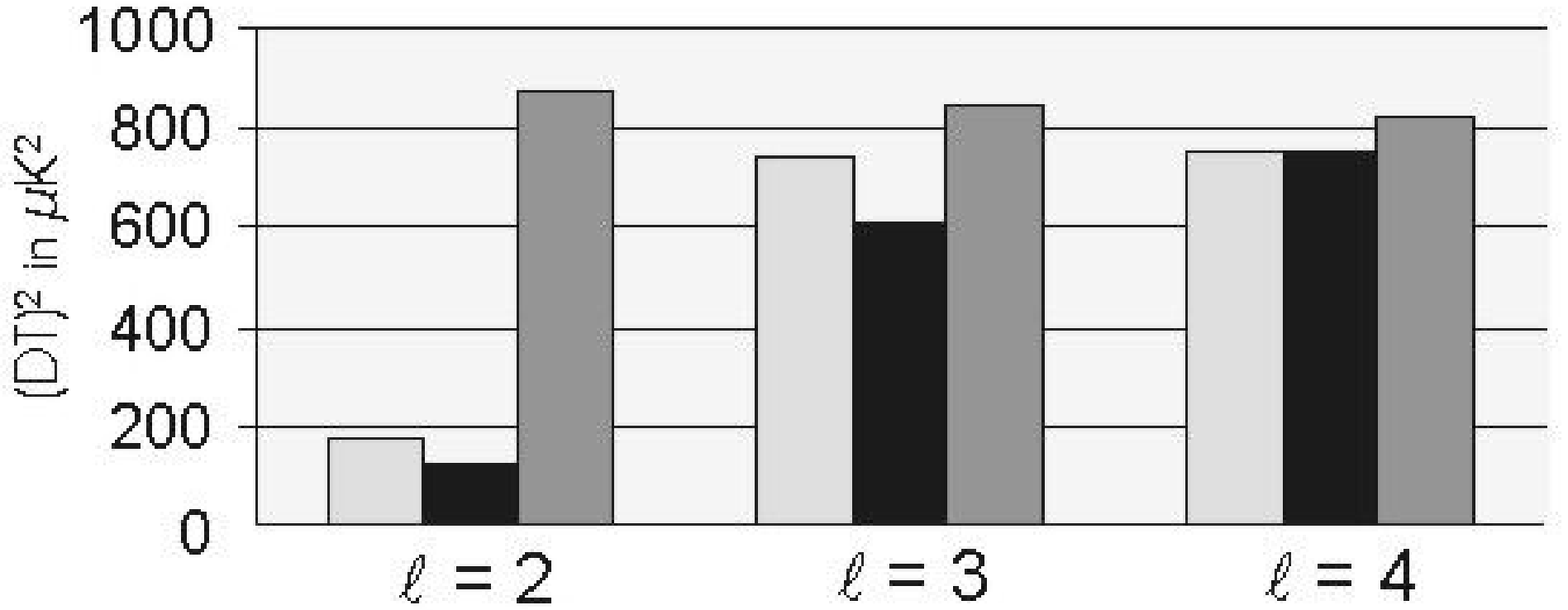}
\caption{Comparison of WMAP power spectrum to Poincar\'e dodecahedral space
and infinite flat universe.  At the low end of the power spectrum WMAP's 
results (drawn in black) match the Poincar\'e dodecahedral space (light grey)
better than they match the expectations for an infinite flat universe (dark grey). 
Computed for a matter energy density parameter $\Omega_m = 0.28$ and a 
cosmological constant energy density parameter $\Omega_{\Lambda} = 0.734$ with Poincar\'e 
dodecahedral space data normalised to the $\ell = 4$ term. \label{fig1}}
\end{figure}

\clearpage 

\begin{figure}
\plotone{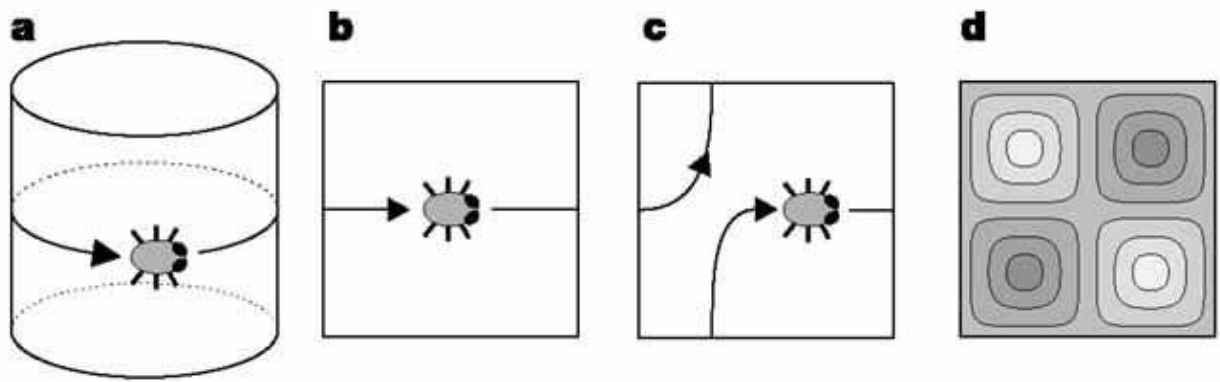}
\caption{Wavelengths of density fluctuations limited by size of finite
``wraparound" universe.  {\bf a)} A 2-dimensional creature living on 
the surface of a cylinder travels due east, eventually going all the way 
around the cylinder and returning to her starting point.  {\bf b)} If we cut 
the cylinder open and flatten it into a square, the creature's path goes out
the square's right side and returns from the left side.  {\bf c)} A flat torus
is like a cylinder, only now the top and bottom sides connect as well as the
left and right. {\bf  d)} Waves in a torus universe may have wavelengths
no longer than the width of the square itself.  To construct a multiconnected 
3-dimensional space, start with a solid polyhedron (for example a cube) 
and identify its faces in pairs, so that any object leaving the polyhedron
through one face returns from the matching face.  Such a multiconnected space
supports standing waves whose exact shape depends on both the geometry 
of the polyhedron and how the faces are identified.  Nevertheless, 
the same principle applies, that the wavelength cannot exceed the size 
of the polyhedron itself.  In particular, the inhabitants of such a space
will observe a cut-off in the wavelengths of density fluctuations.\label{fig2}}
\end{figure}

\clearpage
                
\begin{figure}
\plottwo{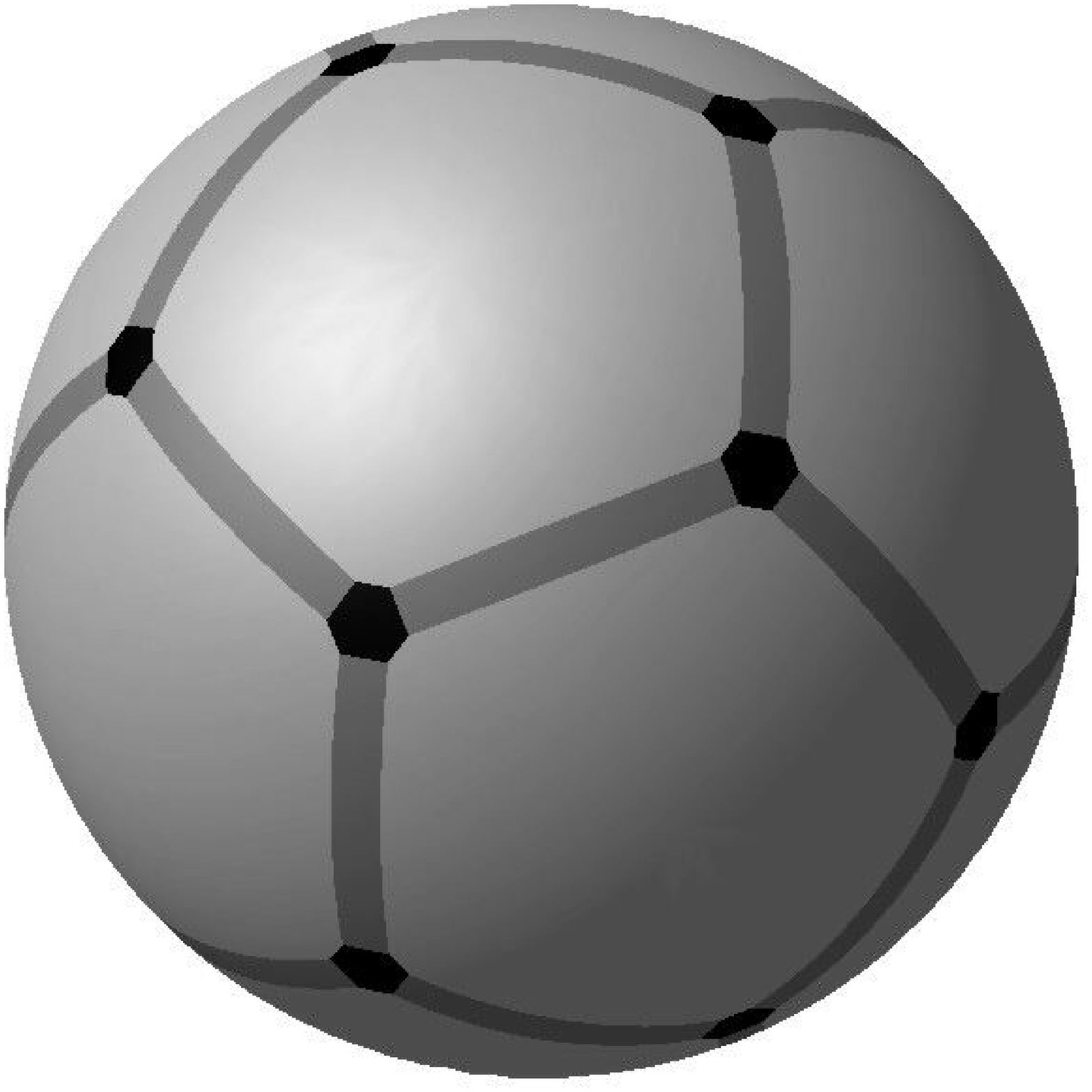}{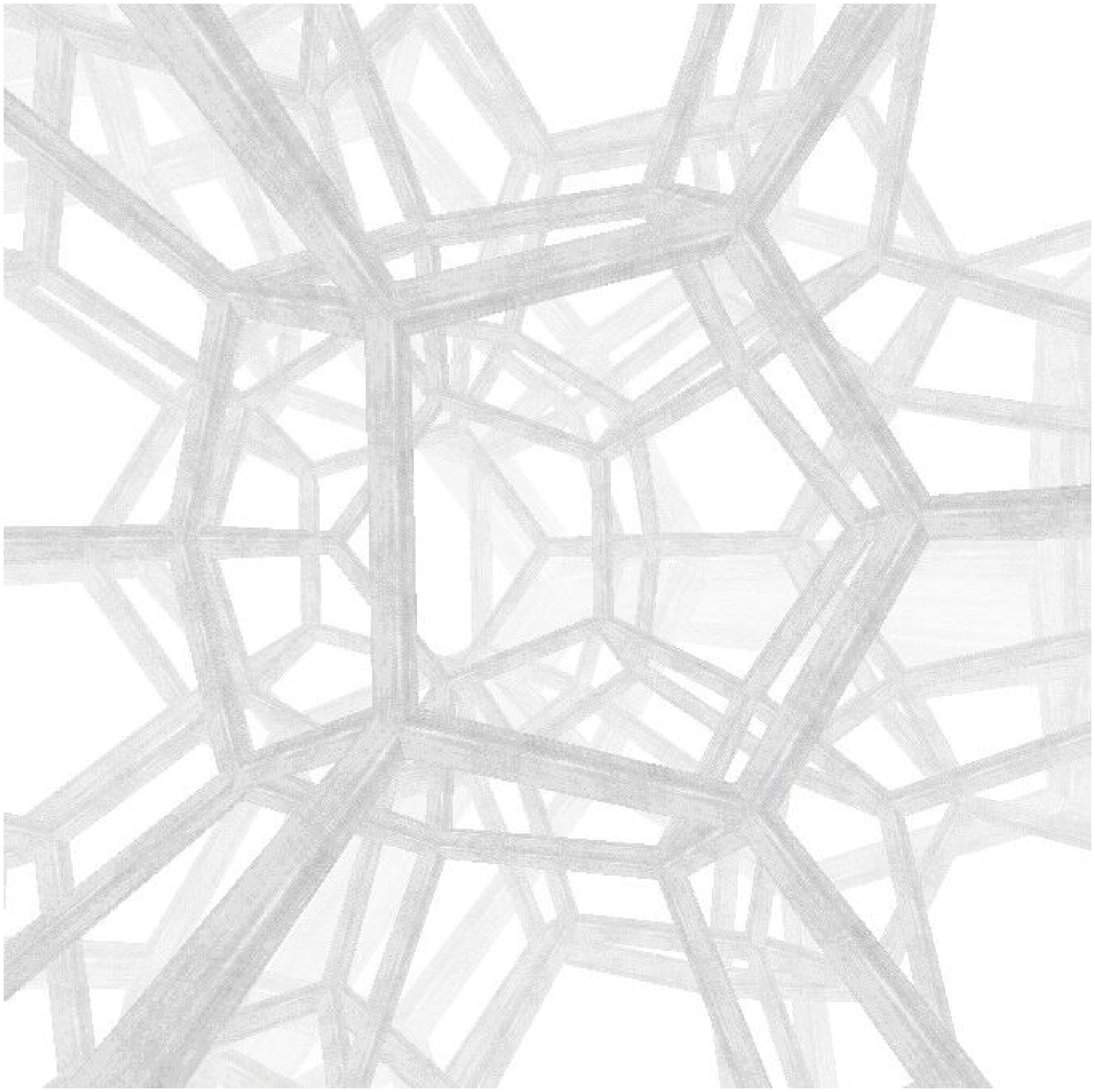}
\caption{Spherical pentagons and dodecahedra fit snugly, unlike their Euclidean
counterparts.  {\bf a)} Twelve spherical pentagons tile the surface of an ordinary
sphere.  They fit together snugly because their corner angles are 
exactly $120^{\circ}$.
Note that each spherical pentagon is just a pentagonal piece of a sphere.
{\bf b)} One hundred twenty spherical dodecahedra tile the surface of a hypersphere.
A hypersphere is the 3-dimensional surface of a 4-dimensional ball.  
Note that each spherical dodecahedron is just a dodecahedral piece of a hypersphere.
The spherical dodecahedra fit together snugly because their edge angles are 
exactly $120^{\circ}$.  In the construction of the Poincar\'e dodecahedral space 
the dodecahedron's 30 edges come together in ten groups of three edges each, 
forcing the dihedral angles to be $120^{\circ}$ and requiring a spherical dodecahedron
rather than a Euclidean one.  Software for visualising spherical dodecahedra
and the Poincar\'e dodecahedral space is available for free download 
from {\tt www.geometrygames.org/CurvedSpaces}.\label{fig3}}
\end{figure}

\clearpage

\begin{figure}
\plotone{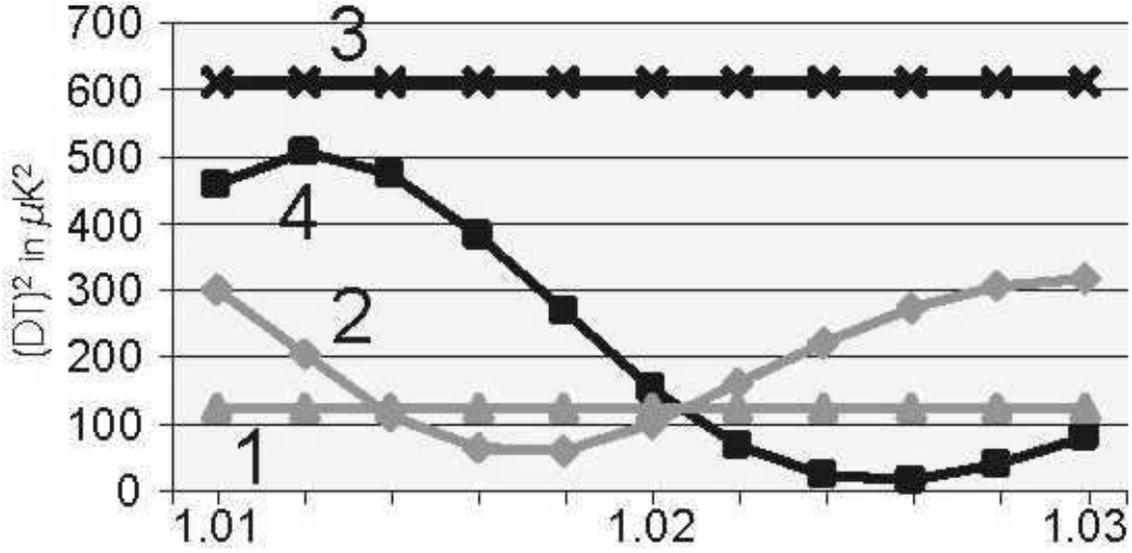}
\caption{Values of the mass-energy density parameter $\Omega_0$ for which the 
Poincar\'e dodecahedral space agrees with WMAP's results.  
The Poincar\'e dodecahedral space quadrupole (trace 2) and octopole 
(trace 4) fit the WMAP
quadrupole (trace 1) and octopole (trace 3) when $1.012 < \Omega_0 < 1.014$.  Larger 
values of $\Omega_0$ 
predict an unrealistically weak octopole.  To obtain these predicted values 
we first computed the Poincar\'e dodecahedral space's eigenmodes using the
Ghost Method of Lehoucq {\it et al.} (2002) with two of the matrix generators computed 
in Appendix B of Gaussmann {\it et al.} (2001), and then applied the method 
of Riazuelo {\it et al.} (2003),
using $\Omega_m = 0.28$ and $\Omega_{\Lambda}= \Omega_0 - \Omega_m$, to obtain 
a power spectrum and to 
simulate sky maps.  Numerical limitations restricted our set of 3-dimensional
eigenmodes to wavenumbers $k < 30$, which in turn restricted the reliable portion
of the power spectrum to $\ell = 2,3,4$.  We set the overall normalisation factor
to match the WMAP data at $\ell = 4$ and then examined the predictions 
for $\ell = 2,3$.
\label{fig4}}
\end{figure}

\end{document}